\documentclass[manuscript, screen]{acmart}
\AtBeginDocument{%
  \providecommand\BibTeX{{%
    \normalfont B\kern-0.5em{\scshape i\kern-0.25em b}\kern-0.8em\TeX}}}
\setcopyright{acmcopyright}
\copyrightyear{2021}
\acmYear{2021}
\makeatletter
\renewcommand\@formatdoi[1]{\ignorespaces}
\makeatother
\acmConference[SimuRec 2021]{SimuRec 2021: Workshop on Simulation Methods for Recommender Systems at ACM RecSys '21}{October 2nd, 2021}{Amsterdam, NL, online}
\acmBooktitle{SimuRec 2021: Workshop on Simulation Methods for Recommender Systems at ACM RecSys 2021, October 2nd, 2021, Amsterdam, NL and online}

\begin{document}
\title{Synthetic Data and Simulators for Recommendation Systems: Current State and Future Directions }

\author{Adam Lesnikowski}
\email{adam.lesnikowski@gmail.com}
\affiliation{%
  \institution{NVIDIA*}
  \authornote{Work for this paper, and paper submission, done while first author at NVIDIA.}
  \city{Redmond}
  \state{Washington}
  \country{USA}
}

\author{Gabriel de Souza Pereira Moreira}
\email{gmoreira@nvidia.com}
\affiliation{%
  \institution{NVIDIA}
  \city{S\~ao Paulo}
  \state{S\~ao Paulo}
  \country{Brazil}
}

\author{Sara Rabhi}
\email{srabhi@nvidia.com}
\affiliation{%
  \institution{NVIDIA}
    \state{Ontario}
  \country{Canada}
}

\author{Karl Byleen-Higley}
\email{karlb@nvidia.com}
\affiliation{%
  \institution{NVIDIA}
  \country{USA}
}

\begin{abstract}
    Synthetic data and simulators have the potential to markedly improve the performance and robustness of recommendation systems. These approaches have already had a beneficial impact in other machine-learning driven fields. We identify and discuss a key trade-off between data fidelity and privacy in the past work on synthetic data and simulators for recommendation systems. For the important use case of predicting algorithm rankings on real data from synthetic data, we provide motivation and current successes versus limitations. Finally we outline a number of exciting future directions for recommendation systems that we believe deserve further attention and work, including mixing real and synthetic data, feedback in dataset generation, robust simulations, and privacy-preserving methods.
\end{abstract}

\maketitle

\section{Introduction}
Synthetic data generation and simulation techniques have been popular and successful in machine learning areas such as computer vision \cite{tremblay2018training} \cite{sankaranarayanan2018learning} and robotics \cite{tobin2018domain} \cite{prakash2019structured}, but have not been broadly explored for recommender systems, with \cite{antulov2014synthetic,slokom2018comparing,slokom2020partially} being notable exceptions. 
These techniques have the potential to address problems such as the lack of publicly available large datasets for research outside of industry. 
This problem is motivated by companies' concerns about user privacy, and the possibility of revealing strategic internal KPIs, such as the level of user engagement and their recurrence in the service, or the growth of the company's item catalog over time.
Synthetic data, when representative of real data, can enable researchers to benchmark, evaluate their methods on datasets of the scale and complexity used in commercial applications. 
The question remains however, to what extent can simulated data effectively balance the trade-off between being close enough to the real data to act as an effective surrogate, while not being close enough to the real data to leak sensitive personal information.
Simulators can potentially generate an infinite amount of synthetic data at very little cost. They bring opportunities for design decisions on recommender systems and algorithms. An algorithm designer could simulate different patterns of user behaviour to evaluate and compare recommendation algorithms, or could emulate the feedback look between users and recommendations. 

The complexity of simulators proposed for recommendation systems varies from simple to complex, depending on its purpose, whether for algorithm comparison, framework development, or simulating the feedback loop between recommendations and user interactions. 
We explore and motivate these uses, and highlight exciting future directions, in this work.

\subsection{Past Successes of Synthetic Data and Simulators}
Domains such as computer vision outside of recommendation systems have benefited markedly from applying machine learning on synthetic or simulated data \cite{sankaranarayanan2018learning}\cite{prakash2019structured}.
Synthetic data in computer visions tasks such as object detection and segmentation has proven to be an effective strategy to increase model validation set performance\cite{tobin2018domain}, and in particular, domain randomization has been surprisingly effective\cite{tremblay2018training}. 

Domain randomization is a technique where simulator parameters are sampled from values which are known to be unrealistic. For example training with domain randomization in a physical simulator would include training with uncommon or extreme values for friction, gravity, and object density. This approach might \emph{a priori} seem detrimental to final model performance by unduly focusing model performance on cases not expected to be encountered in the intended application. Nonetheless this approach of domain randomization has been shown to be effective\cite{tobin2018domain} \cite{tremblay2018training}. 
One explanation for the effectiveness of this technique is that the benefit from a large increase of training diversity outweighs any negative effects from data-distribution shifts towards physically unrealistic scenarios, while another explanation is that the training curriculum is purposefully made more difficult than the intended application, so that the intended application is an easier, more simpler sub-problem than that encountered during the totality of training. In recommendation systems, domain randomization might include setting simulator parameters for users that are known to be unrealistic, such as browsing time, budgets, or spending habits, in order to increase training diversity, so long as validation set performance is improved. 
We believe this past success is a cause of optimism for the use of simulators and synthetic data in recommendation systems.

\section{Fidelity versus Privacy Trade-off}

\subsection{Fidelity}
Synthetic data should share some statistical properties of real data. 
There have been a number of past approaches to capture real data fidelity in synthetic data. For instance \cite{antulov2014synthetic} build synthetic clickstreams through graph walks that explicitly remain faithful to transition probabilities between items and co-occurrence probabilities of items appearing in the same clickstream. The authors measure synthetic dataset fidelity by performance on downstream tasks, such as training recommendation systems on their synthetic data and analyzing the performance of these trained models. 
Past work \cite{belletti2019scalable} evaluate their data-generation method on MovieLens 20M by comparing the item-wise and user-wise rating sums, as well as the singular values, of their generated data matrix versus the real data matrix, in addition to baselines such as histograms of movie ratings between real and generated data.
Past work \cite{lin2020gan} evaluates dataset fidelity of generated time-series data by auto correlation metrics, distribution of generated labels and categorical event types, as well as the prediction of baseline recommendation system algorithm rankings by training and testing on synthetic data.
In general it is important that features have similar distributions, but this requirement is sensitive to the intended usage of synthetic data.
When synthetic data is used to test recommender systems framework and tools based on neural networks, the cardinality and frequency distribution of categorical features is especially important, as they are represented by embeddings in the model. High-cardinality categorical features result in very large embedding tables that may exceed the capacity of a single GPU memory. That poses engineering challenges, like distributing those huge embedding tables to multi-GPU \cite{gupta2020architectural} and minimizing the inter-communication between GPUs by caching the embeddings of popular categorical values \cite{guo2021scalefreectr}, issues addressed by the HugeCTR framework\footnote{https://github.com/NVIDIA/HugeCTR} for example.
On the other hand, these requirements are much more strict for the purpose of comparing different algorithms, as they in general learn patterns from the conditional dependency among the features and the prediction target. Hence ignoring these conditional probabilities between features would be far from a realistic scenario. We discuss this requirement more in Section \ref{sec:alg_ranking} below.

\subsection{Privacy}
\label{sec:privacy_method}
Another research direction of synthetic data generation focuses on using statistical disclosure control techniques to transform original data by hiding specific information. The scale of the resulting data remains the same, but personalized information about user’s preferences is masked to protect his privacy. 
These past works can be classified into two categories: attack modelling \cite{Burke2004IdentifyingAM, Rezaimehr2021} and differential privacy \cite{McSherry2009, Berlioz2015}. 
Attack modelling aims to identify all types of threats, which can take at least three forms: identity disclosure, attribute disclosure and inferential disclosure. Once one or multiple of these threats are identified, synthesized data is generated to prevent attacks. The main limitation of such methods is the requirement of identifying beforehand the attacker’s capabilities and goals, a requirement which is very challenging in practice. In RecSys, previous studies have focused on the classification of recommender attack models \cite{Burke2004IdentifyingAM} and designed simple proof-of-concept models \cite{Slokom2020} to evaluate if the relative performance of algorithms when trained and tested on the synthesized data matches with the relative performance of algorithms trained and tested on the original data. 
By contrast differential privacy aims to prevent attackers from gaining information about their targets, even if the attacker has knowledge about the dataset. One approach towards differential privacy consists of injecting probabilistic noise in the original data while maintaining the same probability distributions. In RecSys differential privacy has been applied to matrix factorization (MF) at different levels of modelling: input perturbation of user-item matrces, private MF optimizers via gradient descent or alternating least squares solvers, and output perturbation. Previous work \cite{Berlioz2015} conducts experiments to evaluate the trade-off between the noise perturbation approaches and MF algorithm accuracy and demonstrated that input perturbation ensures the highest performance. However the authors point out that high degree of noise motivated by ensuring high-level privacy directly impacts the relative ranking of models’ performances. Most recent works are extending differential privacy methods to complex deep recommender systems such as wide and deep architectures \cite{Zhang2021} and collaborative bandits learning \cite{Wang2020}.

\section{Predicting Algorithm Rankings}
\label{sec:alg_ranking}

\subsection{Motivations}
There are very few publicly available high-quality datasets, in terms of size and diversity of available features, likely due to companies concerns on having user privacy or public data that can leak internal company metrics. This scenario limits the research advances in the RecSys field, as scientists outside popular online service companies cannot assess and compare their proposed algorithms on large datasets.
Synthetic generation can be an approach for companies to release data which is similar to its large real data but does not leak this sensitive information, so that third-party researchers can evaluate their proposed algorithms. 

\subsection{Successes and Limitations}
Past works have shown that one can successfully predict what model performance ranking on datasets are, given model performance rankings on simulated and synthetic datasets. For instance \cite{lin2020gan} show a successful prediction of the relative performance rankings of various recommendation systems algorithms trained and tested on real data, obtained by training and testing on synthetic data. Here the algorithm ranking of the authors' proposed GAN-based data-generation method on two different datasets is perfectly aligned, with a correlation ranking of 1.00, with the actual performance among five other algorithms trained and tested on real data. Similarly \cite{slokom2018comparing} provides successful results on the prediction of real algorithm rank orderings from synthetic algorithm rank orderings among three other recommendation system algorithms trained and tested on real data. 
On the other hand, \cite{antulov2014synthetic} provide inconclusive or contradictory evidence that algorithm rank orderings may be successfully predicted for click-stream algorithms, at least for the probabilistic graph walk dataset generation method that the authors propose in this past work.

\section{Future Directions for Simulators and Simulated Data}

\subsection{Data Augmentation, Mixing Synthetic and Real Data for Recommendation Systems}
Data augmentation techniques have been shown to outperform purely synthetic or purely real data in machine learning. 
In computer vision, techniques like random cropping, image mirroring, and color shifting have helped models to generalize better and to achieve improved accuracy\cite{shorten2019survey}. 
Similar strategies have been proposed for raw signals and audio spectograms, such as perturbation and noise injection \cite{ko2015audio, park2019specaugment}, as well as in in NLP \cite{wei2019eda}.
However augmentation techniques for recommender systems have been largely unexplored, with \cite{grbovic2015commerce, wang2019enhancing, wolbitsch2019beggars} being some notable exceptions. 
We believe mixing synthetic and real data can be a promising direction for domains or recommender systems deployments, especially early stages of data collection and small dataset size scenarios.
For research scientists outside large online services companies, it would be very helpful for synthetic data generation to augment real small data, allowing an accurate emulation of algorithm behavior on datasets larger than currently available.

\subsection{Feedback in Dataset Generation}
In production recommendation systems, there is often a back-and-forth process between dataset generation and model training\cite{settles2009active}. In particular a model is likely to be trained on data that a previous model iteration solicited by providing some action that the model selected, like recommending a particular item to a user. This feedback cycle may be positive for performance, in selecting data that future iterations of model training find useful for increasing performance, or this feedback cycle may be pernicious, in either halting or reversing model performance\cite{schmit2018human}. This latter phenomenon may occur by selecting data points which are repetitions in the existing dataset, or more broadly, by focusing on short-term utility rather than promoting dataset coverage or diversity\cite{chaney2018algorithmic}. One concrete example of this latter phenomenon is when a recommendation system recommends a small number of highly popular items, and hence fails to build diverse datasets for future model training iterations. We believe that this back-and-forth process that occurs in commercial applications of recommendation systems, but typically does not in the academic or open-source study of recommendation systems, should be more placed at a higher priority for future recommendation systems research. 

\subsection{Robust Simulations}
Generating high-fidelity synthetic data from real data may not be fully feasible, due partly to biases introduced in the real data by the policies under which the data was collected, and partly due to our imprecise understanding of what dataset properties state-of-the-art models model.
We believe that robust simulators are a promising approach for these concerns. 
For exploring what patterns models are capable of capturing, one can generate synthetic data with known, but not necessarily realistic, properties, towards understanding how the feedback loop between users and algorithms would evolve over time, and which recommendation algorithms would perform better in such scenarios. 
Most public datasets do not include the information necessary to tell apart the actual preferences of the user base and the biasing effects of what the recommender presented to users, which makes unbiased evaluation of new models difficult. 
By evaluating on MovieLens and other public datasets, RecSys as a field has overfit to whatever policy was used during data collection for the MovieLens and other canonical RecSys datasets.
In the robust simulators that we envision, we can make explicit assumptions about what the distribution of true preferences in the user population are, select a particular known logging policies, and generate a set of observations with known properties. 
Simulation allows system administrators to model recommender system dynamics over time.

Simulators can be used to test how generalizable are the proposed recommendation algorithms with respect to edge cases not frequent in real datasets. 
Simulator interpretability allows us to better understand the effect of model parameters we want to test.
One can choose a distribution of true user interests and an observation sampling policy such that we end up with a dataset that has comparable statistics to MovieLens for example, then expand the simulated data to whatever large size is desired.
We do not think RecSys has yet achieved this vision for robust simulators, but simulation frameworks such as RecoGym\cite{rohde2018recogym} and RecSim\cite{ie2019recsim} are promising approaches for this vision.

\subsection{Privacy-Preserving Methods}
Motivated by preserving users sensitive data and global statistics related to business KPIs, large companies have not shared large-scale datasets with external communities. 
Privacy-preserving methods discussed in section \ref{sec:privacy_method} constitute a promise for guaranteeing privacy while releasing large scale datasets for research development. 
However it is still unclear how these noisy aggregated data impact learning effective recommender system models that maintain the relative ranking of different approaches for performance comparison and model selection.  
A recent Criteo challenge organized in collaboration with the CAP21’\footnote{https://medium.com/criteo-engineering/criteo-cap21-privacy-preserving-ai-challenge-9cf9cd880e54} conference aims to benchmark models defined using private constrained training data to explore the trade-off between privacy level and prediction performance. 
In particular individual data is transformed through an embedded, anonymized, and compact representation.
Then machine learning models are trained and tested on two objectives: the privacy attacks protection and the outcome prediction task. 
If such privacy functions are demonstrated to lead to high performance machine learning models, large companies may generate large anonymized synthetic data using a given privacy function, and more openly share it with the RecSys community.

\section{Conclusion}
In this paper, we motivate and state the uses of synthetic data and simulators for recommendation systems. The success of these approaches in other machine learning fields provides promise for these methods. For approaches that use real data, we identify a key trade-off between data fidelity and privacy. The important use case of predicting algorithm rankings using synthetic data is well-motivated, and has had both successes and limitations. Finally there are a number of exciting and promising future directions we believe the field should invest in, including data augmentation that mixes real and synthetic data, feedback in dataset generation in production systems, robust simulators, and privacy-preserving methods.

\bibliographystyle{ACM-Reference-Format}
\bibliography{bib.bib}


\begin{thebibliography}{30}


\ifx \showCODEN    \undefined \def \showCODEN     #1{\unskip}     \fi
\ifx \showDOI      \undefined \def \showDOI       #1{#1}\fi
\ifx \showISBNx    \undefined \def \showISBNx     #1{\unskip}     \fi
\ifx \showISBNxiii \undefined \def \showISBNxiii  #1{\unskip}     \fi
\ifx \showISSN     \undefined \def \showISSN      #1{\unskip}     \fi
\ifx \showLCCN     \undefined \def \showLCCN      #1{\unskip}     \fi
\ifx \shownote     \undefined \def \shownote      #1{#1}          \fi
\ifx \showarticletitle \undefined \def \showarticletitle #1{#1}   \fi
\ifx \showURL      \undefined \def \showURL       {\relax}        \fi
\providecommand\bibfield[2]{#2}
\providecommand\bibinfo[2]{#2}
\providecommand\natexlab[1]{#1}
\providecommand\showeprint[2][]{arXiv:#2}

\bibitem[\protect\citeauthoryear{Antulov-Fantulin, Bo{\v{s}}njak, Zlati{\'c},
  Gr{\v{c}}ar, and {\v{S}}muc}{Antulov-Fantulin et~al\mbox{.}}{2014}]%
        {antulov2014synthetic}
\bibfield{author}{\bibinfo{person}{Nino Antulov-Fantulin},
  \bibinfo{person}{Matko Bo{\v{s}}njak}, \bibinfo{person}{Vinko Zlati{\'c}},
  \bibinfo{person}{Miha Gr{\v{c}}ar}, {and} \bibinfo{person}{Tomislav
  {\v{S}}muc}.} \bibinfo{year}{2014}\natexlab{}.
\newblock \showarticletitle{Synthetic Sequence Generator for Recommender
  Systems--Memory Biased Random Walk on a Sequence Multilayer Network}. In
  \bibinfo{booktitle}{\emph{International Conference on Discovery Science}}.
  Springer, \bibinfo{pages}{25--36}.
\newblock


\bibitem[\protect\citeauthoryear{Belletti, Lakshmanan, Krichene, Chen, and
  Anderson}{Belletti et~al\mbox{.}}{2019}]%
        {belletti2019scalable}
\bibfield{author}{\bibinfo{person}{Francois Belletti}, \bibinfo{person}{Karthik
  Lakshmanan}, \bibinfo{person}{Walid Krichene}, \bibinfo{person}{Yi-Fan Chen},
  {and} \bibinfo{person}{John Anderson}.} \bibinfo{year}{2019}\natexlab{}.
\newblock \showarticletitle{Scalable realistic recommendation datasets through
  fractal expansions}.
\newblock \bibinfo{journal}{\emph{arXiv preprint arXiv:1901.08910}}
  (\bibinfo{year}{2019}).
\newblock


\bibitem[\protect\citeauthoryear{Berlioz, Friedman, Kaafar, Boreli, and
  Berkovsky}{Berlioz et~al\mbox{.}}{2015}]%
        {Berlioz2015}
\bibfield{author}{\bibinfo{person}{Arnaud Berlioz}, \bibinfo{person}{Arik
  Friedman}, \bibinfo{person}{Mohamed~Ali Kaafar}, \bibinfo{person}{Roksana
  Boreli}, {and} \bibinfo{person}{Shlomo Berkovsky}.}
  \bibinfo{year}{2015}\natexlab{}.
\newblock \showarticletitle{Applying Differential Privacy to Matrix
  Factorization} \emph{(\bibinfo{series}{RecSys '15})}.
  \bibinfo{publisher}{Association for Computing Machinery},
  \bibinfo{address}{New York, NY, USA}.
\newblock
\showISBNx{9781450336925}


\bibitem[\protect\citeauthoryear{Burke, Mobasher, Zabicki, and Bhaumik}{Burke
  et~al\mbox{.}}{2004}]%
        {Burke2004IdentifyingAM}
\bibfield{author}{\bibinfo{person}{R. Burke}, \bibinfo{person}{B. Mobasher},
  \bibinfo{person}{Roman Zabicki}, {and} \bibinfo{person}{Runa Bhaumik}.}
  \bibinfo{year}{2004}\natexlab{}.
\newblock \showarticletitle{Identifying Attack Models for Secure
  Recommendation}.
\newblock


\bibitem[\protect\citeauthoryear{Chaney, Stewart, and Engelhardt}{Chaney
  et~al\mbox{.}}{2018}]%
        {chaney2018algorithmic}
\bibfield{author}{\bibinfo{person}{Allison~JB Chaney},
  \bibinfo{person}{Brandon~M Stewart}, {and} \bibinfo{person}{Barbara~E
  Engelhardt}.} \bibinfo{year}{2018}\natexlab{}.
\newblock \showarticletitle{How algorithmic confounding in recommendation
  systems increases homogeneity and decreases utility}. In
  \bibinfo{booktitle}{\emph{Proceedings of the 12th ACM Conference on
  Recommender Systems}}. \bibinfo{pages}{224--232}.
\newblock


\bibitem[\protect\citeauthoryear{Grbovic, Radosavljevic, Djuric, Bhamidipati,
  Savla, Bhagwan, and Sharp}{Grbovic et~al\mbox{.}}{2015}]%
        {grbovic2015commerce}
\bibfield{author}{\bibinfo{person}{Mihajlo Grbovic}, \bibinfo{person}{Vladan
  Radosavljevic}, \bibinfo{person}{Nemanja Djuric}, \bibinfo{person}{Narayan
  Bhamidipati}, \bibinfo{person}{Jaikit Savla}, \bibinfo{person}{Varun
  Bhagwan}, {and} \bibinfo{person}{Doug Sharp}.}
  \bibinfo{year}{2015}\natexlab{}.
\newblock \showarticletitle{E-commerce in your inbox: Product recommendations
  at scale}. In \bibinfo{booktitle}{\emph{Proceedings of the 21th ACM SIGKDD
  international conference on knowledge discovery and data mining}}.
  \bibinfo{pages}{1809--1818}.
\newblock


\bibitem[\protect\citeauthoryear{Guo, Guo, Gao, Tang, He, and Liu}{Guo
  et~al\mbox{.}}{2021}]%
        {guo2021scalefreectr}
\bibfield{author}{\bibinfo{person}{Huifeng Guo}, \bibinfo{person}{Wei Guo},
  \bibinfo{person}{Yong Gao}, \bibinfo{person}{Ruiming Tang},
  \bibinfo{person}{Xiuqiang He}, {and} \bibinfo{person}{Wenzhi Liu}.}
  \bibinfo{year}{2021}\natexlab{}.
\newblock \showarticletitle{ScaleFreeCTR: MixCache-based Distributed Training
  System for CTR Models with Huge Embedding Table}.
\newblock \bibinfo{journal}{\emph{Proceedings of SIGIR'21}}.
\newblock


\bibitem[\protect\citeauthoryear{Gupta, Wu, Wang, Naumov, Reagen, Brooks,
  Cottel, Hazelwood, Hempstead, Jia, et~al\mbox{.}}{Gupta
  et~al\mbox{.}}{2020}]%
        {gupta2020architectural}
\bibfield{author}{\bibinfo{person}{Udit Gupta}, \bibinfo{person}{Carole-Jean
  Wu}, \bibinfo{person}{Xiaodong Wang}, \bibinfo{person}{Maxim Naumov},
  \bibinfo{person}{Brandon Reagen}, \bibinfo{person}{David Brooks},
  \bibinfo{person}{Bradford Cottel}, \bibinfo{person}{Kim Hazelwood},
  \bibinfo{person}{Mark Hempstead}, \bibinfo{person}{Bill Jia},
  {et~al\mbox{.}}} \bibinfo{year}{2020}\natexlab{}.
\newblock \showarticletitle{The architectural implications of facebook's
  dnn-based personalized recommendation}. In \bibinfo{booktitle}{\emph{2020
  IEEE International Symposium on High Performance Computer Architecture
  (HPCA)}}. IEEE, \bibinfo{pages}{488--501}.
\newblock


\bibitem[\protect\citeauthoryear{Ie, Hsu, Mladenov, Jain, Narvekar, Wang, Wu,
  and Boutilier}{Ie et~al\mbox{.}}{2019}]%
        {ie2019recsim}
\bibfield{author}{\bibinfo{person}{Eugene Ie}, \bibinfo{person}{Chih-wei Hsu},
  \bibinfo{person}{Martin Mladenov}, \bibinfo{person}{Vihan Jain},
  \bibinfo{person}{Sanmit Narvekar}, \bibinfo{person}{Jing Wang},
  \bibinfo{person}{Rui Wu}, {and} \bibinfo{person}{Craig Boutilier}.}
  \bibinfo{year}{2019}\natexlab{}.
\newblock \showarticletitle{Recsim: A configurable simulation platform for
  recommender systems}.
\newblock \bibinfo{journal}{\emph{arXiv preprint arXiv:1909.04847}}
  (\bibinfo{year}{2019}).
\newblock


\bibitem[\protect\citeauthoryear{Ko, Peddinti, Povey, and Khudanpur}{Ko
  et~al\mbox{.}}{2015}]%
        {ko2015audio}
\bibfield{author}{\bibinfo{person}{Tom Ko}, \bibinfo{person}{Vijayaditya
  Peddinti}, \bibinfo{person}{Daniel Povey}, {and} \bibinfo{person}{Sanjeev
  Khudanpur}.} \bibinfo{year}{2015}\natexlab{}.
\newblock \showarticletitle{Audio augmentation for speech recognition}. In
  \bibinfo{booktitle}{\emph{Sixteenth annual conference of the international
  speech communication association}}.
\newblock


\bibitem[\protect\citeauthoryear{Lin, Jain, Wang, Fanti, and Sekar}{Lin
  et~al\mbox{.}}{2020}]%
        {lin2020gan}
\bibfield{author}{\bibinfo{person}{Zinan Lin}, \bibinfo{person}{Alankar Jain},
  \bibinfo{person}{Chen Wang}, \bibinfo{person}{Giulia Fanti}, {and}
  \bibinfo{person}{Vyas Sekar}.} \bibinfo{year}{2020}\natexlab{}.
\newblock \showarticletitle{Using GANs for Sharing Networked Time Series Data:
  Challenges, Initial Promise, and Open Questions}. In
  \bibinfo{booktitle}{\emph{Proceedings of the ACM Internet Measurement
  Conference}} (Virtual Event, USA) \emph{(\bibinfo{series}{IMC '20})}.
  \bibinfo{publisher}{Association for Computing Machinery},
  \bibinfo{address}{New York, NY, USA}, \bibinfo{pages}{464–483}.
\newblock
\showISBNx{9781450381383}
\urldef\tempurl%
\url{https://doi.org/10.1145/3419394.3423643}
\showDOI{\tempurl}


\bibitem[\protect\citeauthoryear{McSherry and Mironov}{McSherry and
  Mironov}{2009}]%
        {McSherry2009}
\bibfield{author}{\bibinfo{person}{Frank McSherry} {and} \bibinfo{person}{Ilya
  Mironov}.} \bibinfo{year}{2009}\natexlab{}.
\newblock \showarticletitle{Differentially Private Recommender Systems:
  Building Privacy into the Netflix Prize Contenders}
  \emph{(\bibinfo{series}{KDD '09})}.
\newblock


\bibitem[\protect\citeauthoryear{Park, Chan, Zhang, Chiu, Zoph, Cubuk, and
  Le}{Park et~al\mbox{.}}{2019}]%
        {park2019specaugment}
\bibfield{author}{\bibinfo{person}{Daniel~S Park}, \bibinfo{person}{William
  Chan}, \bibinfo{person}{Yu Zhang}, \bibinfo{person}{Chung-Cheng Chiu},
  \bibinfo{person}{Barret Zoph}, \bibinfo{person}{Ekin~D Cubuk}, {and}
  \bibinfo{person}{Quoc~V Le}.} \bibinfo{year}{2019}\natexlab{}.
\newblock \showarticletitle{Specaugment: A simple data augmentation method for
  automatic speech recognition}.
\newblock \bibinfo{journal}{\emph{arXiv preprint arXiv:1904.08779}}
  (\bibinfo{year}{2019}).
\newblock


\bibitem[\protect\citeauthoryear{Prakash, Boochoon, Brophy, Acuna, Cameracci,
  State, Shapira, and Birchfield}{Prakash et~al\mbox{.}}{2019}]%
        {prakash2019structured}
\bibfield{author}{\bibinfo{person}{Aayush Prakash}, \bibinfo{person}{Shaad
  Boochoon}, \bibinfo{person}{Mark Brophy}, \bibinfo{person}{David Acuna},
  \bibinfo{person}{Eric Cameracci}, \bibinfo{person}{Gavriel State},
  \bibinfo{person}{Omer Shapira}, {and} \bibinfo{person}{Stan Birchfield}.}
  \bibinfo{year}{2019}\natexlab{}.
\newblock \showarticletitle{Structured domain randomization: Bridging the
  reality gap by context-aware synthetic data}. In
  \bibinfo{booktitle}{\emph{2019 International Conference on Robotics and
  Automation (ICRA)}}. IEEE, \bibinfo{pages}{7249--7255}.
\newblock


\bibitem[\protect\citeauthoryear{Rezaimehr and Dadkhah}{Rezaimehr and
  Dadkhah}{2021}]%
        {Rezaimehr2021}
\bibfield{author}{\bibinfo{person}{Fatemeh Rezaimehr} {and}
  \bibinfo{person}{Chitra Dadkhah}.} \bibinfo{year}{2021}\natexlab{}.
\newblock \showarticletitle{A survey of attack detection approaches in
  collaborative filtering recommender systems}.
\newblock \bibinfo{journal}{\emph{Artificial Intelligence Review}}
  (\bibinfo{year}{2021}).
\newblock


\bibitem[\protect\citeauthoryear{Rohde, Bonner, Dunlop, Vasile, and
  Karatzoglou}{Rohde et~al\mbox{.}}{2018}]%
        {rohde2018recogym}
\bibfield{author}{\bibinfo{person}{David Rohde}, \bibinfo{person}{Stephen
  Bonner}, \bibinfo{person}{Travis Dunlop}, \bibinfo{person}{Flavian Vasile},
  {and} \bibinfo{person}{Alexandros Karatzoglou}.}
  \bibinfo{year}{2018}\natexlab{}.
\newblock \showarticletitle{Recogym: A reinforcement learning environment for
  the problem of product recommendation in online advertising}.
\newblock \bibinfo{journal}{\emph{Proceedings of the REVEAL workshop at the
  Twelfth ACM Conference on Recommender Systems (RecSys '18)}}.
\newblock


\bibitem[\protect\citeauthoryear{Sankaranarayanan, Balaji, Jain, Lim, and
  Chellappa}{Sankaranarayanan et~al\mbox{.}}{2018}]%
        {sankaranarayanan2018learning}
\bibfield{author}{\bibinfo{person}{Swami Sankaranarayanan},
  \bibinfo{person}{Yogesh Balaji}, \bibinfo{person}{Arpit Jain},
  \bibinfo{person}{Ser~Nam Lim}, {and} \bibinfo{person}{Rama Chellappa}.}
  \bibinfo{year}{2018}\natexlab{}.
\newblock \showarticletitle{Learning from synthetic data: Addressing domain
  shift for semantic segmentation}. In \bibinfo{booktitle}{\emph{Proceedings of
  the IEEE Conference on Computer Vision and Pattern Recognition}}.
  \bibinfo{pages}{3752--3761}.
\newblock


\bibitem[\protect\citeauthoryear{Schmit and Riquelme}{Schmit and
  Riquelme}{2018}]%
        {schmit2018human}
\bibfield{author}{\bibinfo{person}{Sven Schmit} {and} \bibinfo{person}{Carlos
  Riquelme}.} \bibinfo{year}{2018}\natexlab{}.
\newblock \showarticletitle{Human interaction with recommendation systems}. In
  \bibinfo{booktitle}{\emph{International Conference on Artificial Intelligence
  and Statistics}}. PMLR, \bibinfo{pages}{862--870}.
\newblock


\bibitem[\protect\citeauthoryear{Settles}{Settles}{2009}]%
        {settles2009active}
\bibfield{author}{\bibinfo{person}{Burr Settles}.}
  \bibinfo{year}{2009}\natexlab{}.
\newblock \showarticletitle{Active learning literature survey}.
\newblock  (\bibinfo{year}{2009}).
\newblock


\bibitem[\protect\citeauthoryear{Shorten and Khoshgoftaar}{Shorten and
  Khoshgoftaar}{2019}]%
        {shorten2019survey}
\bibfield{author}{\bibinfo{person}{Connor Shorten} {and}
  \bibinfo{person}{Taghi~M Khoshgoftaar}.} \bibinfo{year}{2019}\natexlab{}.
\newblock \showarticletitle{A survey on image data augmentation for deep
  learning}.
\newblock \bibinfo{journal}{\emph{Journal of Big Data}} \bibinfo{volume}{6},
  \bibinfo{number}{1} (\bibinfo{year}{2019}), \bibinfo{pages}{1--48}.
\newblock


\bibitem[\protect\citeauthoryear{Slokom}{Slokom}{2018}]%
        {slokom2018comparing}
\bibfield{author}{\bibinfo{person}{Manel Slokom}.}
  \bibinfo{year}{2018}\natexlab{}.
\newblock \showarticletitle{Comparing recommender systems using synthetic
  data}. In \bibinfo{booktitle}{\emph{Proceedings of the 12th ACM Conference on
  Recommender Systems}}. \bibinfo{pages}{548--552}.
\newblock


\bibitem[\protect\citeauthoryear{Slokom, Larson, and Hanjalic}{Slokom
  et~al\mbox{.}}{2020a}]%
        {slokom2020partially}
\bibfield{author}{\bibinfo{person}{Manel Slokom}, \bibinfo{person}{Martha
  Larson}, {and} \bibinfo{person}{Alan Hanjalic}.}
  \bibinfo{year}{2020}\natexlab{a}.
\newblock \showarticletitle{Partially Synthetic Data for Recommender Systems:
  Prediction Performance and Preference Hiding}.
\newblock \bibinfo{journal}{\emph{arXiv preprint arXiv:2008.03797}}
  (\bibinfo{year}{2020}).
\newblock


\bibitem[\protect\citeauthoryear{Slokom, Larson, and Hanjalic}{Slokom
  et~al\mbox{.}}{2020b}]%
        {Slokom2020}
\bibfield{author}{\bibinfo{person}{Manel Slokom}, \bibinfo{person}{Martha~A.
  Larson}, {and} \bibinfo{person}{Alan Hanjalic}.}
  \bibinfo{year}{2020}\natexlab{b}.
\newblock \showarticletitle{Partially Synthetic Data for Recommender Systems:
  Prediction Performance and Preference Hiding}.
\newblock \bibinfo{journal}{\emph{CoRR}}  \bibinfo{volume}{abs/2008.03797}
  (\bibinfo{year}{2020}).
\newblock
\showeprint[arxiv]{2008.03797}
\urldef\tempurl%
\url{https://arxiv.org/abs/2008.03797}
\showURL{%
\tempurl}


\bibitem[\protect\citeauthoryear{Tobin, Biewald, Duan, Andrychowicz, Handa,
  Kumar, McGrew, Ray, Schneider, Welinder, et~al\mbox{.}}{Tobin
  et~al\mbox{.}}{2018}]%
        {tobin2018domain}
\bibfield{author}{\bibinfo{person}{Josh Tobin}, \bibinfo{person}{Lukas
  Biewald}, \bibinfo{person}{Rocky Duan}, \bibinfo{person}{Marcin
  Andrychowicz}, \bibinfo{person}{Ankur Handa}, \bibinfo{person}{Vikash Kumar},
  \bibinfo{person}{Bob McGrew}, \bibinfo{person}{Alex Ray},
  \bibinfo{person}{Jonas Schneider}, \bibinfo{person}{Peter Welinder},
  {et~al\mbox{.}}} \bibinfo{year}{2018}\natexlab{}.
\newblock \showarticletitle{Domain randomization and generative models for
  robotic grasping}. In \bibinfo{booktitle}{\emph{2018 IEEE/RSJ International
  Conference on Intelligent Robots and Systems (IROS)}}. IEEE,
  \bibinfo{pages}{3482--3489}.
\newblock


\bibitem[\protect\citeauthoryear{Tremblay, Prakash, Acuna, Brophy, Jampani,
  Anil, To, Cameracci, Boochoon, and Birchfield}{Tremblay
  et~al\mbox{.}}{2018}]%
        {tremblay2018training}
\bibfield{author}{\bibinfo{person}{Jonathan Tremblay}, \bibinfo{person}{Aayush
  Prakash}, \bibinfo{person}{David Acuna}, \bibinfo{person}{Mark Brophy},
  \bibinfo{person}{Varun Jampani}, \bibinfo{person}{Cem Anil},
  \bibinfo{person}{Thang To}, \bibinfo{person}{Eric Cameracci},
  \bibinfo{person}{Shaad Boochoon}, {and} \bibinfo{person}{Stan Birchfield}.}
  \bibinfo{year}{2018}\natexlab{}.
\newblock \showarticletitle{Training deep networks with synthetic data:
  Bridging the reality gap by domain randomization}. In
  \bibinfo{booktitle}{\emph{Proceedings of the IEEE conference on computer
  vision and pattern recognition workshops}}. \bibinfo{pages}{969--977}.
\newblock


\bibitem[\protect\citeauthoryear{Wang, Zhao, Wu, Chopra, Khaitan, and
  Wang}{Wang et~al\mbox{.}}{2020}]%
        {Wang2020}
\bibfield{author}{\bibinfo{person}{Huazheng Wang}, \bibinfo{person}{Qian Zhao},
  \bibinfo{person}{Qingyun Wu}, \bibinfo{person}{Shubham Chopra},
  \bibinfo{person}{Abhinav Khaitan}, {and} \bibinfo{person}{Hongning Wang}.}
  \bibinfo{year}{2020}\natexlab{}.
\newblock \showarticletitle{Global and Local Differential Privacy for
  Collaborative Bandits}. In \bibinfo{booktitle}{\emph{Fourteenth ACM
  Conference on Recommender Systems}} \emph{(\bibinfo{series}{RecSys '20})}.
  \bibinfo{publisher}{Association for Computing Machinery},
  \bibinfo{address}{New York, NY, USA}.
\newblock
\showISBNx{9781450375832}
\urldef\tempurl%
\url{https://doi.org/10.1145/3383313.3412254}
\showURL{%
\tempurl}


\bibitem[\protect\citeauthoryear{Wang, Yin, Wang, Nguyen, Huang, and Cui}{Wang
  et~al\mbox{.}}{2019}]%
        {wang2019enhancing}
\bibfield{author}{\bibinfo{person}{Qinyong Wang}, \bibinfo{person}{Hongzhi
  Yin}, \bibinfo{person}{Hao Wang}, \bibinfo{person}{Quoc Viet~Hung Nguyen},
  \bibinfo{person}{Zi Huang}, {and} \bibinfo{person}{Lizhen Cui}.}
  \bibinfo{year}{2019}\natexlab{}.
\newblock \showarticletitle{Enhancing collaborative filtering with generative
  augmentation}. In \bibinfo{booktitle}{\emph{Proceedings of the 25th ACM
  SIGKDD International Conference on Knowledge Discovery \& Data Mining}}.
  \bibinfo{pages}{548--556}.
\newblock


\bibitem[\protect\citeauthoryear{Wei and Zou}{Wei and Zou}{2019}]%
        {wei2019eda}
\bibfield{author}{\bibinfo{person}{Jason Wei} {and} \bibinfo{person}{Kai Zou}.}
  \bibinfo{year}{2019}\natexlab{}.
\newblock \showarticletitle{Eda: Easy data augmentation techniques for boosting
  performance on text classification tasks}.
\newblock \bibinfo{journal}{\emph{arXiv preprint arXiv:1901.11196}}
  (\bibinfo{year}{2019}).
\newblock


\bibitem[\protect\citeauthoryear{W{\"o}lbitsch, Walk, Goller, and
  Helic}{W{\"o}lbitsch et~al\mbox{.}}{2019}]%
        {wolbitsch2019beggars}
\bibfield{author}{\bibinfo{person}{Matthias W{\"o}lbitsch},
  \bibinfo{person}{Simon Walk}, \bibinfo{person}{Michael Goller}, {and}
  \bibinfo{person}{Denis Helic}.} \bibinfo{year}{2019}\natexlab{}.
\newblock \showarticletitle{Beggars can't be choosers: Augmenting sparse data
  for embedding-based product recommendations in retail stores}. In
  \bibinfo{booktitle}{\emph{Proceedings of the 27th ACM Conference on User
  Modeling, Adaptation and Personalization}}. \bibinfo{pages}{104--112}.
\newblock


\bibitem[\protect\citeauthoryear{Zhang, Mironov, and Hejazinia}{Zhang
  et~al\mbox{.}}{2021}]%
        {Zhang2021}
\bibfield{author}{\bibinfo{person}{Huanyu Zhang}, \bibinfo{person}{Ilya
  Mironov}, {and} \bibinfo{person}{Meisam Hejazinia}.}
  \bibinfo{year}{2021}\natexlab{}.
\newblock \showarticletitle{Wide Network Learning with Differential Privacy}.
\newblock \bibinfo{journal}{\emph{CoRR}}  \bibinfo{volume}{abs/2103.01294}
  (\bibinfo{year}{2021}).
\newblock
\showeprint[arxiv]{2103.01294}
\urldef\tempurl%
\url{https://arxiv.org/abs/2103.01294}
\showURL{%
\tempurl}


\end{thebibliography}

\end{document}